\documentclass[prl,aps,twocolumn,footinbib,showpacs]{revtex4}
\usepackage{bm,graphicx,amsmath,wasysym,txfonts} 
\usepackage{bbm}
\usepackage{epstopdf}
\usepackage{dcolumn}
\usepackage{bm}
\usepackage{graphicx}
\usepackage{amsmath}
\usepackage{latexsym}
\usepackage{amsfonts}
\usepackage{amssymb}
\usepackage{array}
\usepackage{epsfig}
\usepackage{epstopdf}
\newcommand{\ket}[1]{\left\vert#1\right\rangle}
\newcommand{\bra}[1]{\left\langle#1\right\vert}
\newcommand{\nbar}{\overline{n}}

\begin{document}
\title{Using macroscopic entanglement to close the detection loophole in Bell inequality
}

\author{
Youngrong Lim,$^1$  Mauro Paternostro,$^2$  Minsu Kang,$^1$ Jinhyoung Lee,$^3$ and Hyunseok Jeong$^1$
}

\affiliation{
$^1$Center for Macroscopic Quantum Control, Department of Physics and Astronomy, Seoul National University, Seoul, 151-742, Korea\\
$^2$Centre for Theoretical Atomic, Molecular and Optical Physics, School of Mathematics and Physics, Queen's University Belfast, BT7 1NN, UK\\
$^3$Department of Physics, Hanyang University, Seoul 133-791, Korea}
\date{\today}

\begin{abstract}
We consider a Bell-like inequality performed using various instances of multi-photon entangled states to demonstrate that losses occurring {\it after} the unitary transformations used in the nonlocality test can be counteracted by enhancing the ``size" of such entangled states. In turn, this feature can be used to overcome detection inefficiencies affecting the test itself: a slight increase in the size of such states, pushing them towards a more {\it macroscopic} form of entanglement,
significantly improves the state robustness against detection inefficiency, thus easing the closing of the detection loophole. Differently, losses {\it before} the unitary transformations cause decoherence effects that cannot be compensated using macroscopic entanglement.  
\end{abstract}

\maketitle

Quantum mechanics reveals a world quite different from the classical one.
Probably the most surprising consequence of the basic 
assumptions of quantum mechanics is that local realism is no longer tenable. This is proved by violation of Bell's 
inequality~\cite{Bell}. However, the undoubtedly impressive successes in the experimental violation of local realism have not yet reached the stage 
where the violation of Bell's inequality free from well-known experimental loopholes is possible: while the locality loophole can be handled by using ultra-fast analyzers and photonic information carriers, which guarantee the space-like distribution of particles outside the light cone~\cite{weihs}, 
 considerable efforts have been directed towards the closure of the detection loophole through the use of various sorts of quantum states
and measurement schemes~\cite{qudit, atom}. 

While this sets the underlying motivations for our investigation, the following identifies the somehow counterintuitive path that we propose
 to take in order to solve the issue of the detection loophole: we consider an optical setting for the violation of Bell's inequality and argue that only a 
 slight increase in the ``macroscopic nature'' of the two-mode optical resource to be used in the test is sufficient to close the detection loophole.  
Here, ``macroscopic entanglement'' should be intended as the entanglement between macroscopically distinguishable states (we use the term ``macroscopicity'' accordingly)~\cite{note}. The achievement  of macroscopic entanglement is one of the most stimulating topics, at all levels, in modern quantum mechanics in light of the intrinsic interest in the observation of quantum phenomena on a macroscopic scale.
Endeavors in this sense are made difficult  due to the fact that entanglement between macroscopically distinguishable states appears to be possible only under very selective conditions~\cite{atomic,mauro} and it would easily be
destroyed by the interaction with the surrounding world~\cite{zurek}. However, we show that by considering optical states that are very close to experimental realization and are nevertheless endowed with a non-negligible 
macroscopic character, the experimental requirements for the closure of the detection loophole can be significantly lowered.

\begin{figure}[t]
\centerline{\scalebox{0.35}{\includegraphics{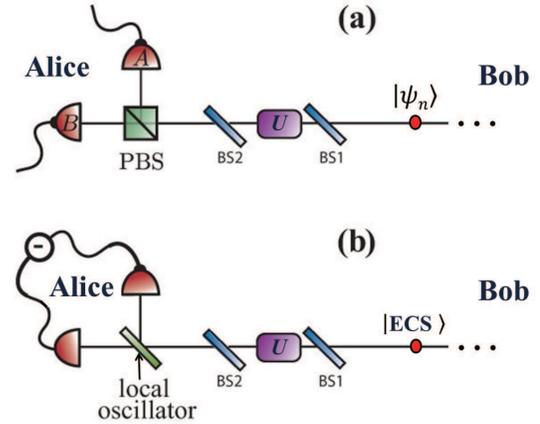}}}
\caption{(Color online)
{\bf (a)} Bell inequality test using entangled polarization state and on/off photodetectors
under photon loss. 
States
$|n_H\rangle$ and $|n_V\rangle$
are discriminated by Alice using
a polarization beam splitter (PBS) and two photodetectors $A$ and $B$.
A local unitary operation, $U$, is required when performing a Bell inequality test.
Photon loss is modeled by
a beam splitter (BS) with transmittivity $\eta$. BS1 and BS2 are employed to consider losses both before and after the unitary operation, respectively.
Bob undergoes the same procedure with his chosen unitary operation.
{\bf (b)} Analogous 
set-up for entangled coherent state, $|{\rm ECS}\rangle$,  and homodyne detection. Two component states, $|\alpha\rangle$
and $|-\alpha\rangle$, of an entangled coherent state are distinguished by homodyne detection.}
\label{schemi}
\end{figure}

We first focus on entanglement between photon number states
\begin{equation}
|\psi_n\rangle=\frac{1}{\sqrt{2}}(|n_H\rangle|n_V\rangle+|n_V\rangle|n_H\rangle),
\label{ens}
\end{equation}
where $|n_H\rangle$ ($|n_V\rangle$) denotes a state of $n$ photons, all with horizontal (vertical) polarization in a single spatial mode. 
State~\eqref{ens} is equivalent to a GHZ-type entangled  state
$|H\rangle^{\otimes 2n}+|V\rangle^{\otimes 2n}$~\cite{Gao2010}: it is obvious that local unitary operations and spatial mode rearrangement can convert one to the other. 
These types of states have been experimentally demonstrated for values of $n$ up to $5$ with fidelity 0.56 ($n=4$ with 0.78)~\cite{Walther2004,Gao2010}. The components
 $|n_H\rangle$ and $|n_V\rangle$ of Eq.~(\ref{ens}) can be considered macroscopically distinguishable when $n$ is large.

Suppose now that, experimentally, we only have extremely inefficient detectors, such as human eyes, so that    
only ``macroscopic difference'' can be noticed. Notice that such a possibility has been recently considered in the context of experimental revelation of entanglement and non-locality in entangled states involving macroscopic 
components~\cite{sticatzi,main,Stobinska2011} as well as quantum-to-classical transition \cite{main,Buzek1995,brukner08,Rae2011}.
If $n$ is large enough, the two states can still be discriminated using two 
inefficient detectors, $A$ and $B$, and a polarization beam splitter (PBS) as shown in Fig. 1.
If only detector $A$ ($B$) clicks, one can tell that the input state was $|n_H\rangle$ ($|n_V\rangle$).
The measurement scheme fails only when the relevant detector misses all the incoming photons.
Assuming that efficiency of each detector is $\eta$,
the success probability $P_s$ for the scheme is 
\begin{equation}
P_s=1-(1-\eta)^n,
\end{equation}
which can be made arbitrarily close to unity by increasing $n$, for any non-zero value of $\eta$.  
Distinguishability between the two states can be made arbitrarily good by
increasing the photon number.

Let us introduce the measurement operator 
\begin{equation}
\hat{O}=\sum_{k=1}^n\big(|k_H\rangle\langle k_H|-|k_V\rangle\langle k_V|\big)+|0\rangle\langle 0|,
\label{op}
\end{equation}
whose last term is necessary to include a ``no-click'' event at both the $A$ and $B$ photodetectors in Fig.~\ref{schemi}{\bf (a)}, a case
that should be included in a loophole-free Bell test.
The correlation function is constructed as
\begin{equation}
E^p(\theta_a,\theta_b,\eta){=}
{\rm Tr}_{aba^\prime b^\prime}[\hat{O}_a \otimes \hat{O}_b |\Psi_n(\theta_a,\theta_b,\eta)\rangle
\langle\Psi_n(\theta_a,\theta_b,\eta)|]
\label{correl}
\end{equation}
where the superscript $p$ is used to indicate that polarization-entangled states are used and 
\begin{equation}
\label{stato}
|\Psi_n(\theta_a,\theta_b,\eta)\rangle{=}\hat{\cal B}^p_{a a^\prime}(\eta,\theta_a)\otimes
\hat{\cal B}^p_{b b^\prime}(\eta,\theta_b)|\psi_n\rangle_{ab}|00\rangle_{a' b^\prime}.
\end{equation}
We have introduced  
$\hat{\cal B}^p_{j j^\prime}(\eta,\theta_j)=\hat{B}_{j j^\prime}(\eta)\hat{U}^p(\theta_j)$~$(j=a,b)$ 
with ${\hat B}_{ab}(\eta)=e^{\frac{\zeta}{2}({\hat a}^\dagger {\hat b}
-{\hat a} {\hat b}^\dagger)}$ the operator of a beam splitter of transmittivity $\eta{=}(\cos\zeta)^2$ and $\hat{U}^p(\theta_j)=\exp[i\theta_j(|n_H\rangle_j\langle n_V|+h.c.)]$ a rotation about the $x$-axis of the Bloch sphere of a polarization qubit $\{|n_H\rangle,|n_V\rangle\}_j$ 
encoded in mode $j$ ($\forall{n}{\in}{\mathbb N}$).
As such unitary operation depends on the photon number $n$, it needs the
non-linear Hamiltonian 
$\hat H_n =g( {{\hat a^n_H}}{\hat a_V^{\dag n}}e^{i\phi} + h.c.)$
to be realized: by choosing the interaction time, any value of $\theta_j$ can be obtained. 
One can in principle implement this type of highly non-linear Hamiltonian
by decomposing it into series of Gaussian unitaries and
cubic operations~\cite{Seckin,Petr}.

\begin{figure}[t]
\centerline{\scalebox{0.45}{\includegraphics[width=550pt]{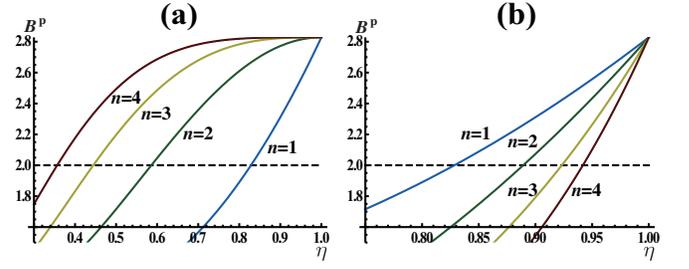}}}
\caption{(Color online) The optimal Bell function of the entangled polarization state against $\eta$ for two cases depicted in Fig.1{\bf (a)}.
(a) The case of photon losses after the unitary operation that is equivalent to inefficient detection. As photon number $n$ increases,
nonlocality shown by Bell-inequality violations gets more robust against photon losses.
The detection-inefficiency threshold decreases to 35.6$\%$ for $n=4$.
(b) The case of photon losses before the unitary operation. }
\label{polar}
\end{figure}

Using the correlation function (\ref{correl}),
it is straightforward to construct the Bell function~\cite{CHSH} $B^p(\theta_a,\theta_b,\theta'_a,\theta'_b,\eta)=E^p(\theta_a,\theta_b,\eta)+E^p(\theta_a,\theta_b^\prime,\eta)
+E^p(\theta_a^\prime,\theta_b,\eta)-E^p(\theta_a^\prime,\theta_b^\prime,\eta)$,
which should satisfy $|B^p|\leq 2$ under the assumptions of local realism. In our case, the correlation function is
\begin{equation}
E^p({\theta _a},{\theta _b},\eta ){=}(1-\eta )^{2 n}-[1-(1-\eta )^n]^2 \cos\left[2 \left(\theta _a+\theta _b\right)\right].
\end{equation}
In Fig.~\ref{polar}{\bf (a)} we have plotted the optimized Bell function $|B^p|_{\rm max}$ against detection efficiency $\eta$ for several values of
$n$. For $n=1$, the efficiency of minimum violation (detection-inefficiency threshold) is 82.8$\%$, which
is a well known value for the maximally entangled qubit states~\cite{82}.
As it can be clearly seen, when $n$ grows (hence increasing the `macroscopic' nature of the entangled state at hand), the detection-inefficiency threshold decreases, thus showing that low efficiencies can be compensated by an enhanced macroscopic character.

We now address an important point.
As we are interested in the effects of low-efficiency detectors, 
we have so far considered losses after the unitary rotations
and prior to the detection, {\it i.e.}, only BS2 was considered  in Fig.~\ref{schemi}{\bf (a)}. 
What happens if losses occur before the unitary operations as modeled by
BS1 in Fig.~\ref{schemi}{\bf (a)}?  
In order to assess this case, we only need to replace $\hat{\cal B}_{jj'}(\eta,\theta_j)$ in Eq.~\eqref{stato}
with $\hat{\cal B}^{p}_{j j^\prime}(\eta,\theta_j)=\hat{U}^{p}(\theta_j)\hat{B}_{j j^\prime}(\eta)$~$(j{=}a,b)$. 
In this case [see Fig.~\ref{polar}{\bf (b)}], the configuration with increasing photon number becomes very fragile against photon losses.
This is due to the fact that losses before the unitary operations must be treated as decoherence rather than detection inefficiency.
As the effective decoherence rate is faster for larger $n$, simply increasing it and using local unitary operations cannot cure the corresponding spoiling effects. For more plausible result we need to consider both effect at once. This is the reason why we do not indicate
the case of higher $n$ despite the fact that $n=4$ is not sufficiently {\it macroscopic}. More discussions will be presented at the end of the manuscript.

As a second significant example, we consider the entangled coherent state (ECS)~\cite{sanders}
\begin{equation}
\ket{\text{ECS}}=N(|\alpha,\alpha\rangle+|-\alpha,-\alpha\rangle),
\end{equation}
where $|{\pm}\alpha\rangle$ is a coherent state of amplitude $\pm\alpha\in{\mathbb C}$ and 
$N{=}[2(1 + {e^{ - 4{{\left| \alpha  \right|}^2}}})]^{-1/2}$  is a normalization factor.
Such states can be generated using a 50:50 beam splitter and coherent-state superpositions, $|\alpha\rangle+|-\alpha\rangle$, which have been experimentally demonstrated
\cite{cat-gen}. Recently, a nonlocal generation of the entangled coherent state
was successfully  demonstrated \cite{ecs-gen}.

The set-up that we consider in this case is illustrated in Fig.~\ref{schemi}{\bf (b)} but BS1 is ignored for now.
Similarly to our previous example, we take unitary operations $\hat{U}^{ecs}_{jj'}(\theta_j)~(j{=}a,b)$ embodied by effective rotation performed in the space spanned by the basis $\{\left| \alpha  \right\rangle, \left| -\alpha  \right\rangle\}$. For large values of $|\alpha|$, such transformations can be performed approximately using a properly arranged cascade of single-mode Kerr-like nonlinearities and displacement operations~\cite{paternostro10,SJR,main}. A Bell test can then be constructed by using homodyne measurements, whose outcomes are dichotomized in a way that a logical outcome $+1$ ($-1$) is associated to positive (negative) expectation values of the in-phase quadrature operator of each mode. In Refs.~\cite{main,SJR,paternostro10,mckeown10,mckeown11}, some of us have proven that this approach can be successfully applied to the exploration of non-classicality tests, including multipartite non-locality and quantum contextuality.
\begin{figure}[t]
\centerline{\scalebox{0.42}{\includegraphics[width=400pt]{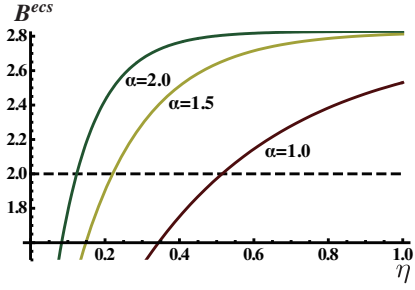}}}
\caption{(Color online)
The optimized Bell parameter for an ECS. Similarly to polarization-entangled states, by increasing the macroscopic character of  the state (given in this case by the amplitude $\alpha$), the Bell parameter becomes more robust for photon losses. For $\alpha$=1, the detection-inefficiency threshold is about $50\%$.}
\label{ECS}
\end{figure}

Here, we model inefficient homodyne detectors in a way fully analogous to what has been done previously for photodetectors [cf. Fig.~\ref{schemi}{\bf (b)}]. For simplicity and without limitations to the validity of our analysis, in what follows we assume $\alpha\in{\mathbb R}$. We study the behavior of the optimized Bell parameter $|B^{ecs}(\theta_a,\theta_b,\theta'_a,\theta'_b,\eta)|$, which has been constructed using the correlation functions ${E}^{ecs}(\theta_a,\theta_b,\eta)=P_{++}+P_{--}-P_{+-}-P_{-+}$, where we have defined the probabilities
\begin{equation}
{P}_{kl}{=}\!\!\int^{k_s}_{k_i}\!\!\!dx\!\!\int^{l_s}_{l_i}\!\!\!dy\langle{x_ax_b}|{\rm Tr}_{a'b'}[|\Phi(\theta_a,\theta_b,\eta)\rangle\langle\Phi(\theta_a,\theta_b,\eta)|]|x_ax_b\rangle
\end{equation}
with $|\Phi(\theta_a,\theta_b,\eta_{a^\prime b^\prime})\rangle{=}\hat{\cal B}^{ecs}_{aa^\prime}(\eta,\theta_a)\otimes\hat{\cal B}^{ecs}_{bb^\prime}(\eta,\theta_b)|\text{ECS}\rangle_{ab}|00\rangle_{a'b'}$, $\hat{\cal B}^{ecs}_{j j^\prime}(\eta,\theta_j){=}\hat{B}_{j j^\prime}(\eta)\hat{U}^{ecs}(\theta_j)$~$(j{=}a,b)$, the subscripts $k,l{=}\pm$ that correspond to the assigned measurement outcomes $\pm{1}$ and the integration limits $+_s=\infty,~+_i=-_s=0$ and $-_i=-\infty$. Moreover, $|x_{a}\rangle$ ($|x_{b}\rangle$) is an  eigenstate of the quadrature operator $\hat{x}_a=\hat{a}+\hat{a}^\dag$ ($\hat{x}_b=\hat{b}+\hat{b}^\dag$) with eigenvalue $x_a$ ($x_b$).

In Fig.~\ref{ECS} we plot the Bell parameter against the homodyne detector efficiency $\eta$ for various choices of $\alpha$.
It is clear in this case too that by increasing the macroscopic character of the state (i.e. by increasing $\alpha$), we gain in robustness and
the violation of the Bell inequality is possible even for large losses. For instance, at $\alpha=1$, violation is 
possible for $50\%$ detection efficiencies, which sets an important benchmark. 
Comparing with the previous case, ECS is more robust to detection inefficiency than polarization entangled state under same average photon
number. For example, the detection-inefficiency threshold is about 0.83 for the polarization entangled state of two photons, but it is about 0.5 for the ECS.
Moreover, by addressing the case of photon losses occurring before the local unitary operations, modeled by BS1 in Fig.~\ref{schemi}{\bf (b)}, one can also see that an opposite effect, very much similar to the one observed for polarization entangled state, is in order: larger values $\alpha$ imply a larger detection-inefficiency threshold for the violation of the Bell-CHSH inequality. This can be easily understood and verified by studying the Markovian master equation
$\partial_t\varrho=\gamma\hat{a}\varrho\hat{a}^\dag-\frac{\gamma}{2}
(\hat{a}^\dag\hat{a}\varrho+\varrho\hat{a}^\dag\hat{a})$ 
with $\varrho$ the density matrix of a boson with associated annihilation 
(creation) operator $\hat{a}$ ($\hat{a}^\dag$), $\gamma$ the loss rate and $t$ the interaction time. 
Decoherence described by this maser equation is equivalent to the beam splitter loss:
the relation between the beam splitter loss $\eta$ before the unitary operation and the loss rate $\gamma$
in the master equation is given by $\eta=\exp[-\gamma t]$.

We now extend our analysis to an entangled thermal state (ETS)  \cite{jr06}
\begin{equation}
\label{ruotato}
\begin{split}
&\rho_{ab}^{ets}(\theta_a,\theta_b){=}N_+\!\!\int\!\!\int{d}^2\alpha{d}^2
\beta{P}^{th}_{\alpha}(V,d){P}^{th}_{\beta}(V,d)\ket{\text{ecs}}_{ab}\!\bra{\text{ecs}}
\end{split}
\end{equation}
with $\ket{\rm ecs}=|{\alpha,\beta}\rangle+|{-\alpha,-\beta}\rangle$, $N_\pm=[2(1\pm{e^{-{4d^2}/{V}}}/{V^2})]^{-1}$ and $P^{th}_{\alpha}(V,d)=\frac{2}{\pi(V-1)}e^{-\frac{2|\alpha-d|^2}{(V-1)}}$ the Gaussian thermal distribution with variance $V=2(\nbar-d^2)+1$ ($\nbar$ is the mean photon number) and center $d$ (with respect to the origin of the phase space). This state can be created by entangling two single-mode thermal states mutually displaced by $d$ and has been used to prove the possibility to violate local realism with coarse grained homodyne measurements and thermal local states~\cite{main}.   

We subject Eq.~\eqref{ruotato} to the local rotations $\hat{U}^{ecs}_{jj'}(\theta_j)$ used in order to assess an ECS and eventually arrive at the following expression for the correlation function $C^{ets}(\theta_a,\theta_b)$ that enters in the Bell parameter $B^{ets}(\theta_a,\theta_b,\theta_a^\prime,\theta_b^\prime)=
C^{ets}(\theta_a,\theta_b)+C^{ets}(\theta_a^\prime,\theta_b)+C^{ets}(\theta_a,\theta_b^\prime)-C^{ets}(\theta_a^\prime,\theta_b^\prime)$
\begin{widetext}
\begin{equation}
\begin{split}
C^{ets}(\theta_a,\theta_b)&=
V_1V_2\left\{e^{4i\theta_a}g_\eta(\theta_a)
\left[
Qg_\eta(\theta_b)s(\theta_b)
+ie^{\frac{2d^2}{V}}V_1h(\theta_b)
\left(f_{-,\eta}(\theta_b)-e^{8i\theta_b}f_{+,\eta}(\theta_b)\right)\right]\right.\\
&\left.+V_1h(\theta_a)\left[ie^{2\theta_b(2i+\frac{V\theta_b}{d^2})}g_{\eta}(\theta_b)
s(\theta_b)\left(f_-(\theta_a)-e^{8i\theta_b}f_+(\theta_a)\right)
+4V_1h(\theta_b)\left(e^{8i\theta_a}f_{-,\eta}(\theta_b)f_{+,\eta}(\theta_a)
+e^{8i\theta_a}f_{-,\eta}(\theta_a)f_{+,\eta}(\theta_b)\right)
\right]\right\}
\end{split}
\end{equation}
\end{widetext}
where $s(\theta_i)={\rm sign(\theta_i)}$ and
\begin{equation}
\begin{split}
&h(\theta_i)=e^{\frac{2(d^4+\theta_i^2)}{d^2V}},~g_\eta(\theta_j){=}{\rm Erfi}\left[\frac{\sqrt{2}\eta\theta_j}{d\sqrt{V^2-\eta^2V(V-1)}}\right],\\
&V_1=(1/8)(1+V^2e^{\frac{4d^2}{V}})^{-1},~V_2=e^{{-4i(\theta_a+\theta_b)-\frac{2(1+V^2)(\theta_a^2+\theta_b^2)}{d^2V}}},\\ 
&Q=8e^{{4i\theta_b+\frac{2V(\theta_a^2+\theta_b^2)}{d^2}}},~f_{\pm,\eta}(\theta_j){=}{\rm Erf}\left[\frac{\sqrt{2}\eta(d^2\pm i V\theta_j)}{d\sqrt{1+\eta^2(V-1)}}\right].
\end{split}
\end{equation}
\begin{figure}[ht]
{\bf (a)}\hspace{4cm}{\bf (b)}
{\scalebox{0.44}{\includegraphics{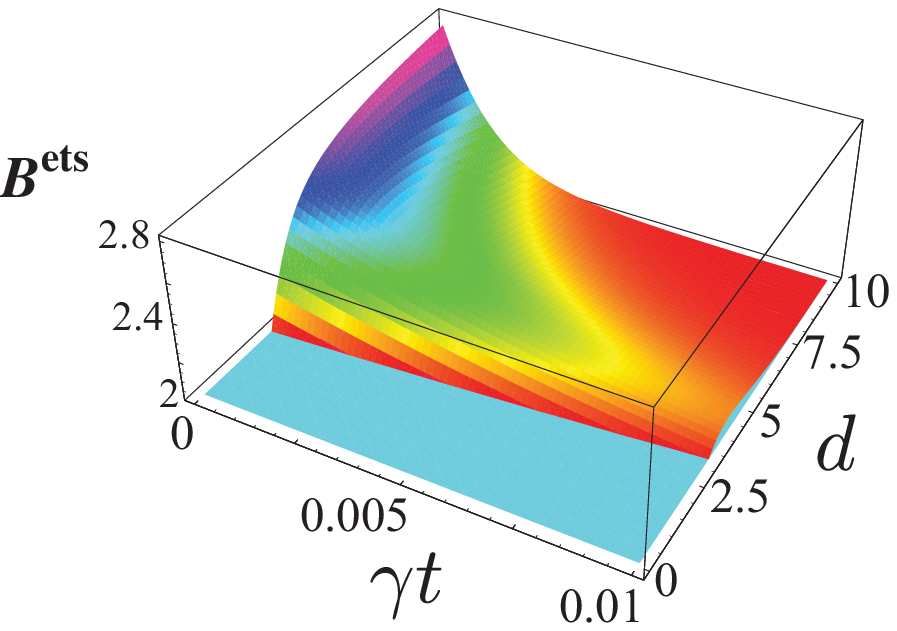}}}~~~~\scalebox{0.43}{\includegraphics{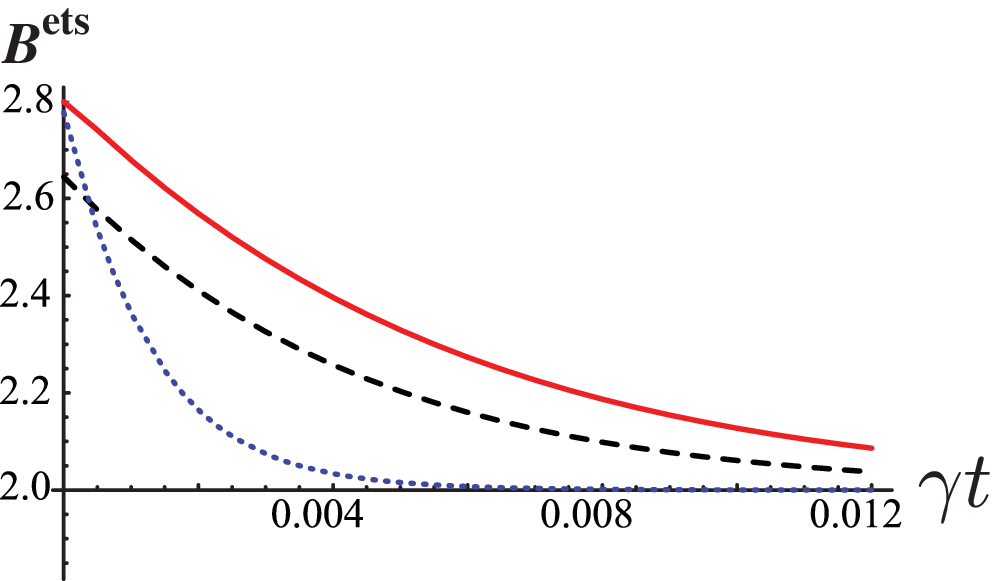}}
\caption{(Color online) {\bf (a)} The optimized Bell function against the dimensionless decoherence parameter $\gamma t$ and $d$ 
for $V=10$. The floor of the figure shows the limit for local realistic theories. {\bf (b)}
The optimized Bell function 
against $\gamma t$
for $V=1.001$, $d=5$ (solid curve), 
$V=10$, $d=5$ (dashed curve) and $V=10$, $d=10$ (dotted curve).
}
\label{fig:dec} 
\end{figure}
While a detailed study of the behavior of $B^{ets}$ against
$V$ and $d$ is provided in Ref.~\cite{main}, here we mention that the effects that detection inefficiencies have 
on the violation of the Bell-CHSH inequality can indeed be compensated for a large violation of Bell's inequality 
by increasing the displacement amplitude 
$d$.  
This means that photon losses at detectors can be overcome
by increasing the amplitude $d$ of the initial thermal state.
However, photon losses {\it before} the measurements at the detectors
(for example, during the generation process of the ETS), which will
cause decoherence of the state itself, may reduce entanglement 
and destroy the violation of Bell-CHSH inequality. This is what we now ascertain.

We can exploit the formulation of the lossy evolution of a bosonic system given by Phoenix~\cite{phoenix} to find out that 
the correlation function to use for the evaluation of the Bell parameter of an ETS affected by losses at a rate $\gamma$
can be explicitly calculated, although the analytic expression turns out to be quite lengthy. In Fig.~\ref{fig:dec}{\bf (a)}, we show the behavior of the Bell parameter against the separation $d$ between the thermal components 
and the dimensionless dissipation parameter $\gamma{t}$ for $V=10$ (arbitrary choice). Evidently, already for
  modest values of $d$ an ETS violates the Bell-CHSH inequality for any value of the decoherence parameter.
 Larger values of $V$ simply require a larger threshold in $d$ to show violation of the Bell-CHSH inequality,
 which is therefore still a property of our genuinely macroscopic states affected by decoherence.
 In panel {\bf (b)} we have compared the Bell functions obtained
assigning the value of $V$ and $d$. We see that the increase of $V$ does not boost the decrease of $B^{ets}$ with $\gamma{t}$.
Indeed, the slopes of the curves in Fig.~\ref{fig:dec}{\bf (b)}
of an ETS with $V=10$ and $d=5$ (dashed curve) and of a (nearly) pure ECS with $V=1.001$ and $d=5$ 
(solid curve) are very close to each other. Clearly,
a {larger separation in phase space} (i.e. larger $d$) causes a quicker destruction of the violation of the Bell-CHSH inequality.

\begin{figure}[t!]
{\scalebox{0.35}{\includegraphics{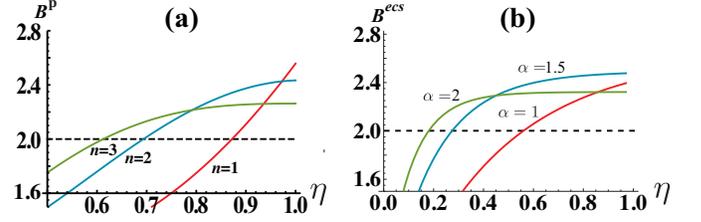}}}
\caption{(Color online) {\bf (a)} The optimal Bell function for three values of $n$ in an entangled polarization state plotted against $\eta$
 when 5\% losses occur before the unitary operation. {\bf (b)} The analogous quantity calculated for an ECS having $\alpha=1,1.5$ and $2$ and subjected to $15\%$ losses.}
\label{fig:both}
\end{figure}

We finally consider a more realistic situation where losses before the unitary operation is present together with detection inefficiency.
We include both the effects in our calculations by using beam splitter operations before {\it and} after the unitary operation as
$\hat{B}_{j j^\prime}(\eta_2)\hat{U}(\theta_j)\hat{B}_{j j^{\prime\prime}}(\eta_1)$,
where $\eta_1$ ($\eta_2$) is the parameter determining losses before the unitary operation (the effects of detection efficiency).
The results in Fig.~\ref{fig:both}{\bf (a)} show that the entangled polarization state of $n=3$ causes the required detection efficiency to be larger than 61\% for 5\% losses before the unitary operation.
As shown in Fig.~\ref{fig:both}{\bf (b)}, the ECS is found to be significantly more efficient: at $\alpha=2$, an ECS shows Bell violations when detection efficiency is larger than $\simeq17\%$ for 15\% losses. Considering that about 250m length for traveling photons would be sufficient to be free from the locality loophole \cite{weihs}, this range of values is not far from experimental reality when using telecom fibers~\cite{Gisin02}.

We discussed several examples of entangled multi-photon states to show that ``macroscopic quantum correlations"
 may indeed be used in order to overcome limitations in fundamental tests of physics performed with inefficient
 detectors: the detection loophole can be made less relevant in Bell tests conducted by using states involving
a large number of excitations and, as such, verging towards classicality.
In contrast to this, losses occurring before the local operations needed to run a Bell-like test affect the quality of the test itself in a way that cannot be corrected by simply considering a more ``macroscopic" resource.  
Our results reveal a previously unknown relation between 
macroscopic entanglement and measurement efficiency in Bell inequality tests.
They reinforce in an unexpected way the idea that quantum features are indeed observable at larger, quasi-macroscopic scales.
As we have shown, state macroscopicity  can in fact be used to magnify
such features and ascertain them in a more reliable way.

 The authors thank M. S. Kim for stimulating discussions. This work was supported by the NRF grant funded by the Korea government (MEST) (No. 3348-20100018), the World Class University (WCU) program and the UK EPSRC.


\begin{thebibliography} {99}


\bibitem{Bell} J. S. Bell, Physics {\bf 1}, 195 (1964).

\bibitem{weihs}  G. Weihs, T. Jennewein, C. Simon, H. Weinfurter, and A. Zeilinger, Phys. Rev. Lett. {\bf 81}, 5039 (1998).

\bibitem{atom} D. N. Matsukevich, P. Maunz, D. L. Moehring, S. Olmschenk, and C. Monroe, Phys. Rev. Lett. {\bf 100}, 150404 (2008).

\bibitem{qudit} T. V\'{e}rtesi, S. Pironio, and N. Brunner, Phys. Rev. Lett. {\bf 104}, 060401 (2010).

\bibitem{atomic} J. Hald, J. L. S\o rensen, C. Schori, and E. S. Polzik, Phys. Rev. Lett. {\bf 83}, 1319 (1999).

\bibitem{note} The word ``macroscopic'' in this paper is used in a relative sense rather than defining a boundary between macroscopic and microscopic states.

\bibitem{mauro} M. Paternostro, D. Vitali, S. Gigan, M. S. Kim, C. Brukner, J. Eisert, and M. Aspelmeyer, Phys. Rev. Lett. {\bf 99}, 250401 (2007).

\bibitem{zurek} W. H. Zurek, Phys. Today {\bf 44}, 36 (1991).



\bibitem{Walther2004} 
P. Walther, J.-W. Pan, M. Aspelmeyer, R. Ursin, S. Gasparoni, and A. Zeilinger, 
Nature {\bf 429}, 158 (2004).

\bibitem{Gao2010} 
W.-B. Gao,  C.-Y. Lu, X.-C. Yao, P. Xu, O. G\"{u}hne, A. Goebel, Y.-A. Chen, C.-Z. Peng, Z.-B. Chen, and J.-W. Pan,
Nature Phys. {\bf 6}, 331 (2010).


\bibitem{sticatzi} F. De Martini, F. Sciarrino, and C. Vitelli, Phys. Rev. Lett. {\bf 100}, 253601 (2008);
P. Sekatski, N. Brunner, C. Branciard, N. Gisin, and C. Simon,
Phys. Rev. Lett. {\bf 103}, 113601 (2009); E. Pomarico {\it et al.}, arXiv:1104.2212; 
N. Spagnolo, C. Vitelli, M. Paternostro, F. De Martini, and F. Sciarrino,
Phys. Rev. A {\bf 84}, 032102 (2011).

\bibitem{main} H. Jeong, M. Paternostro, T. C. Ralph, Phys. Rev. Lett. {\bf 102}, 060403 (2009).

\bibitem{Stobinska2011} M. Stobi\'nska, P. Sekatski, A. Buraczewski, N. Gisin, G. Leuchs, Phys. Rev. A {\bf 84}, 034104 (2011).

\bibitem{Buzek1995} V. Bu\v{z}ek, M. S. Kim, and M. G. Kim, J. Kor. Phys. Soc. {\bf 28}, 123 (1995).

\bibitem{brukner08}  J. Kofler and \v{C}. Brukner, Phys. Rev. Lett. {\bf 99}, 180403 (2007); Phys. Rev. Lett. {\bf 101}, 090403 (2008).


\bibitem{Rae2011} S. Raeisi, P. Sekatski, and C. Simon, Phys. Rev. Lett. {\bf 107}, 250401 (2011).

\bibitem{Seckin}
S. Sefi and P. van Loock, Phys. Rev. Lett. {\bf 107}, 170501 (2011).

\bibitem{Petr}
P. Marek, R. Filip, and A. Furusawa, Phys. Rev. A {\bf 84}, 053802 (2011).

\bibitem{CHSH} J. F. Clauser, M. A. Horne, A. Shimony, and R. A. Holt,
Phys. Rev. Lett. {\bf 23},
880 (1969).

\bibitem{82} A. Garg and N. D. Mermin, Phys. Rev. D {\bf 35}, 3831 (1987).

\bibitem{sanders} B. C. Sanders, Phys. Rev. A {\bf 45}, 6811 (1992).

\bibitem{cat-gen} 
A. Ourjoumtsev, H. Jeong, R. Tualle-Brouri, and Ph. Grangier, Nature (London) \textbf{448}, 784 (2007);
H. Takahashi {\it et al.}, Phys. Rev. Lett. \textbf{101}, 233605 (2008).

\bibitem{ecs-gen}
A. Ourjoumtsev, F. Ferreyrol, R. Tualle-Brouri, and P. Grangier,
Nature Physics {\bf 5}, 189 (2009).

\bibitem{SJR} M. Stobi\'nska, H. Jeong, and T. C. Ralph,
Phys. Rev. A {\bf 75}, 052105 (2007).

\bibitem{paternostro10} M. Paternostro and H. Jeong, Phys. Rev. A {\bf 81}, 032115 (2010).

\bibitem{mckeown10} G. McKeown, F.L. Semiao, H. Jeong, and M. Paternostro, Phys. Rev. A {\bf 82},  022315  (2010); C.-W. Lee, M. Paternostro, and H. Jeong, Phys. Rev. A {\bf 83}, 022102 (2011).

\bibitem{mckeown11} G. McKeown, M.G.A. Paris, and M. Paternostro, Phys. Rev. A {\bf 83}, 062105 (2011).

\bibitem{jr06} H. Jeong and T.C. Ralph, Phys.Rev.Lett. {\bf 97}, 100401 (2006);
Phys.Rev.A {\bf 76}, 042103 (2007).


\bibitem{phoenix} S. J. D. Phoenix, Phys. Rev. A {\bf 41}, 5132 (1990).

\bibitem{Gisin02} N. Gisin, G. Ribordy, W. Tittel, and H. Zbinden, Rev.
Mod. Phys. {\bf 74}, 145 (2002).







\end{thebibliography}
\end{document}